\documentclass{vienna-conf2019}
\usepackage{graphicx}
\usepackage{hyperref}
\usepackage[]{natbib}  
\usepackage{epstopdf}

\def\BibTeX{{\rm B\kern-.05em{\sc i\kern-.025em b}\kern-.08em
    T\kern-.1667em\lower.7ex\hbox{E}\kern-.125emX}}
\bibpunct{(}{)}{;}{a}{}{,}  

\begin{document}

\TitreGlobal{Stars and their variability observed from space}


\title{New fully evolutionary models for asteroseismology of
ultra-massive white dwarf stars}

\runningtitle{Asteroseismology of ultra-massive WD stars}

\author{A. H. C\'orsico$^{1,}$}\address{Facultad de Ciencias Astron\'omicas y Geof\'isicas, Universidad Nacional de La Plata, Paseo delBosque s/n, (1900), La Plata, Argentina}\address{Instituto de Astrof\'isica de La Plata, IALP, CONICET-UNLP, La Plata, Argentina}

\author{F. C. De Ger\'onimo$^{1,2}$}

\author{M. E. Camisassa$^{1,2}$}

\author{L. G. Althaus$^{1,2}$}

\setcounter{page}{237}


\maketitle


\begin{abstract}
  Ultra-massive hydrogen-rich (DA spectral type) white dwarf (WD)
  stars ($M_{\star} > 1M_{\odot}$) coming from single-star
  evolution are expected to harbor cores made of $^{16}$O and $^{20}$Ne,
  resulting from semi-degenerate carbon burning when the progenitor
  star evolves through the super asymptotic giant branch (S-AGB)
  phase. These stars are expected to be crystallized by the time they
  reach the ZZ Ceti instability strip ($T_{\rm eff} \sim 12\,500$ K).
  Theoretical models predict that crystallization leads to a separation
  of $^{16}$O and $^{20}$Ne in the core of ultra-massive WDs, which impacts their
  pulsational properties. This property offers a unique opportunity to
  study the processes of crystallization. Here, we present the first
  results of a detailed asteroseismic analysis of the best-studied
  ultra-massive ZZ Ceti star BPM~37093. As a second step, we plan to
  repeat this analysis using ultra-massive DA WD models with C/O cores
  in order to study the possibility of elucidating the core chemical
  composition of BPM~37093 and shed some light on its possible
  evolutionary origin. We also plan to extend this kind of analyses to
  other stars observed from the ground and also from space missions
  like Kepler and TESS.
\end{abstract}



\section{Input physics, evolution/pulsation codes, and stellar models}

\begin{figure}[ht!]
 \centering
 \includegraphics[width=0.5\textwidth,clip]{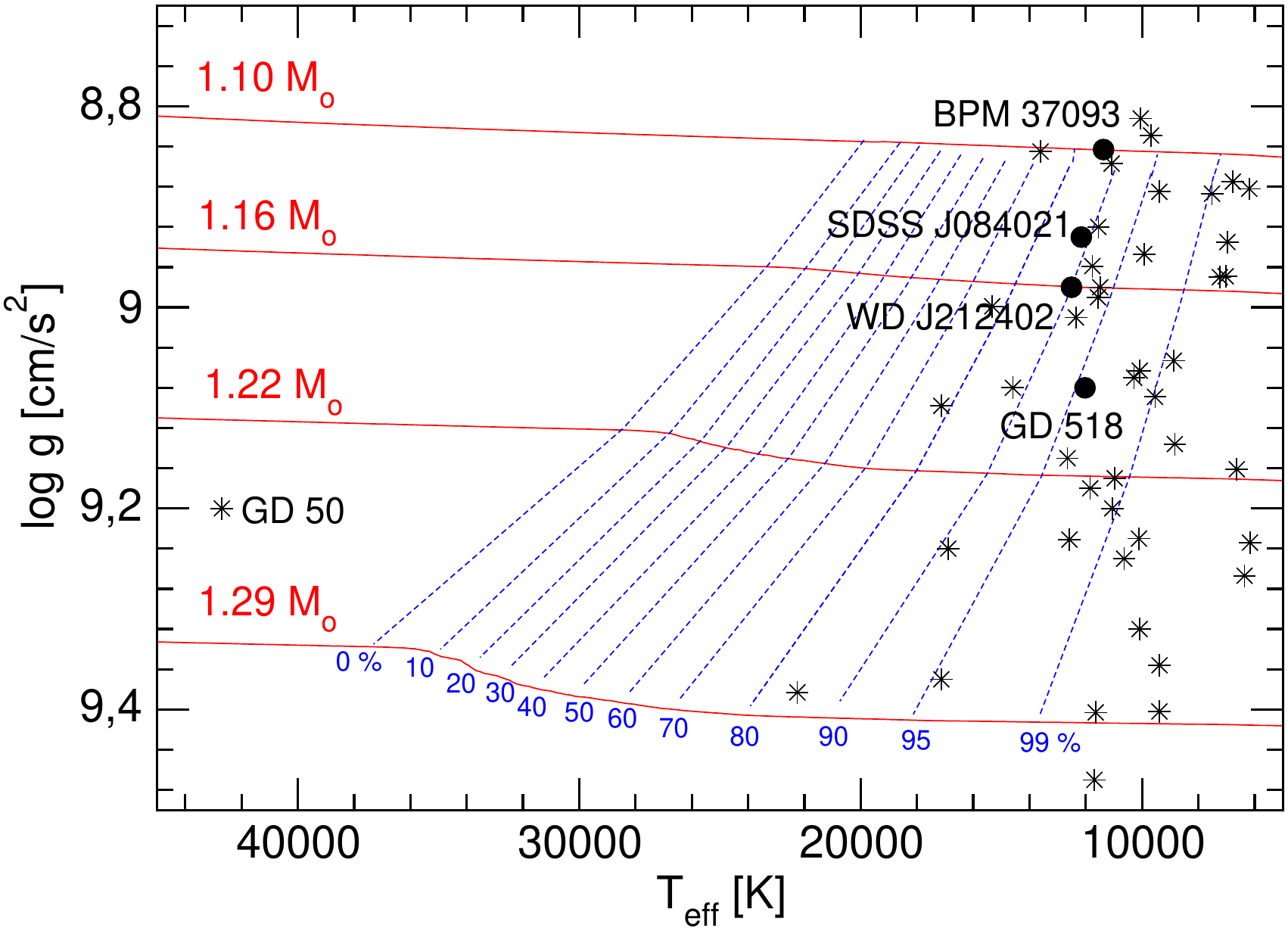}      
 \caption{Evolutionary tracks (red solid lines) of the ultra-massive
   DA WD models computed by \cite{2019A&A...625A..87C}
   in the $T_{\rm eff}- \log g$ plane.
   Blue dashed lines indicate 0, 10, 20, 30, 40, 50, 60, 70, 80, 90,
   95 and 99 \% of crystallized mass. The location of ultra-massive
   DA WD stars are indicated with black star symbols, and the
   ultra-massive ZZ Ceti stars are emphasized with black circles.}
  \label{corsico:fig1}
\end{figure}

The evolutionary models were generated by \cite{2019A&A...625A..87C}
employing the {\tt LPCODE} evolutionary code. The evolutionary tracks are
shown in Fig. \ref{corsico:fig1}. The input physics of {\tt LPCODE} is
described in  \cite{2019A&A...625A..87C} and we refer the interested
reader to that paper for details. Of particular importance in this work is the
treatment of crystallization. Cool WD stars are supposed to
crystallize due to the strong Coulomb interactions in their very dense
interior \citep{1968ApJ...151..227V}. In our models,
crystallization sets in when the energy of the Coulomb interaction
between neighboring ions is much higher than their thermal energy.
The release of latent heat, and the
release of gravitational energy associated with changes in the
chemical composition profile induced by crystallization, are
consistently taken into account. The chemical redistribution due to
phase separation and the associated release of energy have been
considered following \cite{2010ApJ...719..612A}, appropriately modified by
\citep{1968ApJ...151..227V} for ONe plasmas. To assess the enhancement of
$^{20}$Ne in the crystallized core, we used the azeotropic-type phase
diagram of \cite{2010PhRvE..81c6107M}.  The pulsation code used to compute
nonradial $g$(gravity)-mode pulsations is the adiabatic version of the
{\tt LP-PUL} pulsation code described in \cite{2006A&A...454..863C}. To
account for the effects of crystallization on the pulsation spectrum
of g modes, we adopted the ``hard sphere'' boundary conditions
\citep[see][]{1999ApJ...526..976M}.  Our ultra-massive WD models have stellar
masses $M_{\star}= 1.10, 1.16, 1.22$, and $1.29 M_{\odot}$. They result from the
complete evolution of the progenitor stars through the S-AGB
phase. The core and inter- shell chemical profiles of our models at
the start of the WD cooling phase were obtained from \cite{2010A&A...512A..10S}.

\section{Chemical profiles and the Brunt-V\"ais\"al\"a frequency}

The cores of our models are composed mostly of $^{16}$O and $^{20}$Ne
and smaller amounts of $^{12}$C, $^{23}$Na, and $^{24}$Mg. Since
element diffusion and gravitational settling operate throughout the WD
evolution, our models develop pure H envelopes. The He content of our
WD sequences is given by the evolutionary history of progenitor star,
but instead, the H content of our canonical envelopes $[\log(M_{\rm H}/M_{\star}=
-6]$ has been set by imposing that the further evolution does not
lead to H thermonuclear flashes on the WD cooling track. We have
expanded our grid of models by artificially generating new sequences
with thinner H envelopes $[\log(M_{\rm H}/M_{\star}= -7, -8, -9, -10]$,
for each stellar mass value. This artificial procedure has been done at
high-luminosity stages of the WD evolution. The temporal changes of
the chemical abundances due to element diffusion are assessed by using
a new full-implicit treatment for time-dependent element diffusion
\citep{2020A&A...633A..20A}.

\begin{figure}[ht!]
 \centering
 \includegraphics[width=0.42\textwidth,clip]{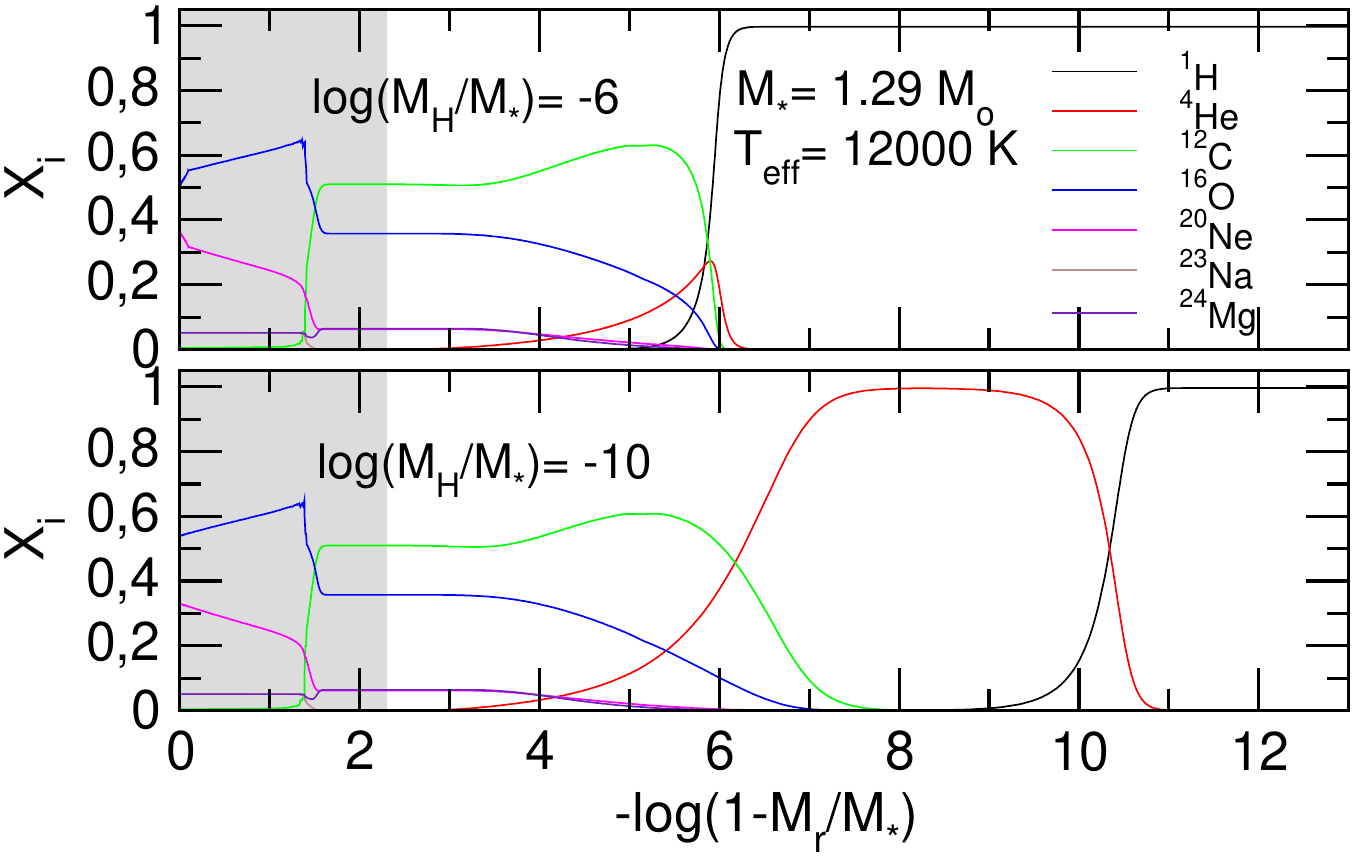}%
 \includegraphics[width=0.53\textwidth,clip]{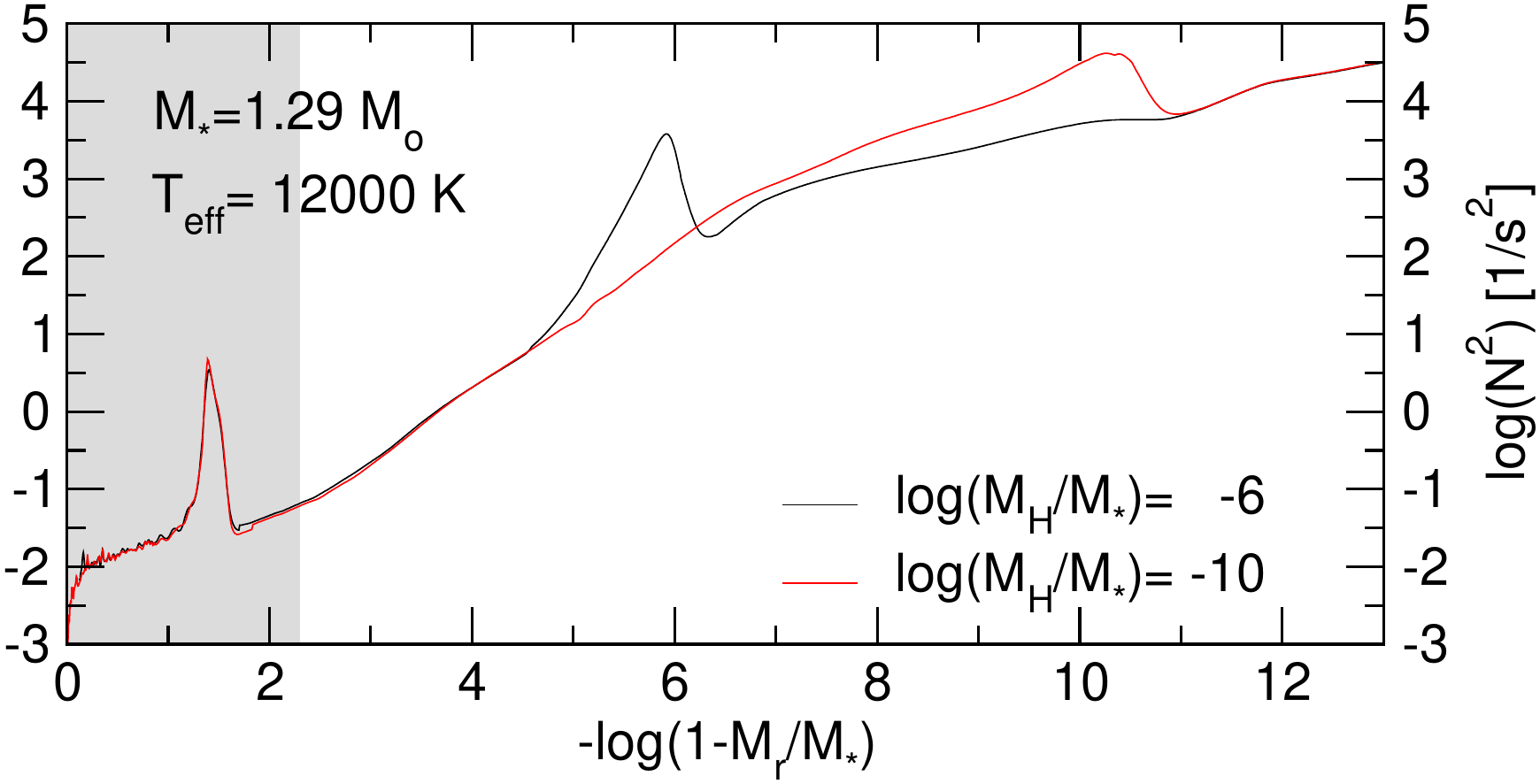}      
 \caption{{\bf Left:} Abundances by mass of the different chemical
   species as a function of the fractional mass, corresponding to
   ONe-core WD models with $M_{\star}= 1.29 M_{\odot}$, $T_{\rm eff} \sim  12\,000$ K
   and two different H envelope thicknesses, as indicated. The
   percentage of crystallized mass fraction of the models is 99.5 \%
   (gray region). {\bf Right:} Logarithm of the squared Brunt-V\"ais\"al\"a
frequency, corresponding to the same ONe-core WD
models with $M_{\star}= 1.29 M_{\odot}$, $T_{\rm eff} \sim  12\,000$ K
and $\log(M_{\rm H}/M_{\star})= -6$, and $-10$ shown in the left panel.}
  \label{corsico:fig2}
\end{figure}

The chemical profiles in terms of the fractional mass for $1.29 M_{\star}$
ONe-core WD models at $T_{\rm eff} \sim 12\,000$ K and H envelope
thicknesses $\log(M_{\rm H}/M_{\star})= -6$ and $-10$ are shown in
the left panel of Fig. \ref{corsico:fig2}. A pure He
buffer develops as we consider thinner H envelopes.
At this effective temperature, the chemical
rehomogeneization due to crystallization has already finished, giving
rise to a core where the abundance of $^{16}$O ($^{20}$Ne) increases
(decreases) outward. In the right panel of Fig. \ref{corsico:fig2}
we show the logarithm of the squared Brunt-V\"ais\"al\"a
frequency corresponding to the same models shown in the left panel of the
figure. The peak at $-\log(1-M_r/M_{\star}) \sim 1.4$, which is due
to the abrupt step at the triple chemical transition between
$^{12}$C, $^{16}$O, and $^{20}$Ne, is within the solid
part of the core, so it has no relevance for the mode-trapping
properties of the models. This is because, according to the
hard-sphere boundary conditions adopted for the pulsations, the
eigenfunctions of $g$ modes do not penetrate the crystallized region.
In this way, the mode trapping properties are entirely determined by
the presence of the He/H transition, which is located in more external
regions for thinner H envelopes. The pulsation properties of these
models have been explored by \cite{2019A&A...621A.100D} and
\cite{2019A&A...632A.119C}.

\begin{table}[]
\centering
\caption{Frequencies and periods of BPM~37093 \citep{2004ApJ...605L.133M},
  along with the theoretical periods, harmonic degrees, radial
orders, and period differences of our best-fit model.}
\begin{tabular}{cc|cccc}
\hline
\hline
$\Pi^{\rm O}$  & $\nu$       & $\Pi^{\rm T}$  & $\ell$ & $k$ & $\delta_i$\\ 
$[$sec$]$ & [$\mu$Hz] &  [sec]         &        &     &  [sec] \\ 
\hline
511.7 & 1954.1 & 512.4 & 2 & 29 & $-0.7$\\
531.1 & 1882.9 & 531.9 & 1 & 17 & $-0.8$\\
548.4 & 1823.5 & 548.1 & 2 & 31 & $0.3$\\ 
564.1 & 1772.7 & 565.3 & 2 & 32 & $-1.2$\\
582.0 & 1718.2 & 583.0 & 2 & 33 & $-1.0$\\
600.7 & 1664.9 & 599.9 & 2 & 34 & $0.8$\\
613.5 & 1629.9 & 613.8 & 1 & 20 & $-0.3$\\
635.1 & 1574.6 & 632.2 & 2 & 36 & $2.9$\\
\hline
\end{tabular}
\label{table:1}
\end{table}

\begin{table}
\centering
\caption{The main characteristics of BPM 37093}
\begin{tabular}{l|cc}
\hline
\hline
Quantity                     & Spectroscopy                     & Asteroseismology             \\
\hline
$T_{\rm eff}$ [K]            & $11\,370\pm 500$       & $11\,650\pm 40$   \\
$M_{\star}/M_{\odot}$         & $1.098\pm 0.1$          & $1.16 \pm 0.014$    \\ 
$\log g$ [cm/s$^2$]          & $8.843 \pm 0.05$        & $8.970 \pm 0.025 $  \\ 
$\log (L_{\star}/L_{\odot})$  & ---                              & $-3.25\pm 0.01$ \\  
$\log(R_{\star}/R_{\odot})$   & ---                              & $-2.234\pm 0.006$ \\  
$\log(M_{\rm H}/M_{\star})$  & ---                              & $-6 \pm 0.26$   \\  
$\log(M_{\rm He}/M_{\star})$  & ---                             & $-3.8$ \\           
$M_{\rm cr}/M_{\star}$    &   $0.935$                              & $0.923$ \\  
$X_{^{16}{\rm O}}$ cent.               & ---                               &  $0.52$ \\     
$X_{^{20}{\rm Ne}}$ cent.               &  ---                            & $0.34$  \\
\hline
\hline
Quantity                 &  Measured                & Asteroseismology \\  
\hline
$\overline{\Delta \Pi}_{\ell= 1}$ [s]  &---                             & 29.70  \\    
$\overline{\Delta \Pi}_{\ell= 2}$ [s]  &  $17.3\pm0.9$                 & 17.63 \\    
\hline
\hline
Quantity                     & Astrometry (\emph{Gaia})                    & Asteroseismology             \\
\hline
$d$  [pc]                     & $14.81$                            &     $11.32$       \\ 
$\pi$ [mas]                   & $67.5$                           &        $88.3$       \\ 
\hline
\hline
\end{tabular}
\label{table:2}
\end{table}

\section{Application: asteroseismological analysis of the ultra-massive
  ZZ Ceti star BPM~37093}

BPM~37093 is the first ultra-massive ZZ Ceti star discovered by
\cite{1992ApJ...390L..89K}. This star is characterized by $T_{\rm
  eff}= 11\,370$ K and $\log g= 8.843$ \citep{2016IAUFM..29B.493N}. We
searched for a pulsation model that best matches the individual
pulsation periods of BPM~37093. The goodness of the match between the
theoretical pulsation periods ($\Pi_k^{\rm T}$) and the observed
periods ($\Pi_i^{\rm O}$) is measured by means of a merit function
defined as $\chi^2(M_{\star}, M_{\rm   H}, T_{\rm   eff})=
\frac{1}{N} \sum_{i=1}^{N} \min[(\Pi_i^{\rm   O}-\Pi_k^{\rm T})^2]$,
where $N$ is the number of observed periods. The WD model
that shows the lowest value of  $\chi^2$, if exists, is adopted as the
``best-fit model''. We assumed two possibilities for the mode
identification: (i) that all of the observed periods correspond to $g$
modes with $\ell= 1$, and (ii) that the observed periods correspond to a
mix of $g$ modes with $\ell= 1$ and $\ell= 2$. We considered the eight
periods employed by \cite{2004ApJ...605L.133M} (see Table \ref{table:1}).
The case (i) did not show clear solutions compatible with BPM~37093
in relation to its spectroscopically-derived effective temperature.
Instead, the case (ii) resulted in a clear seismological solution for
a WD model with $M_{\star}= 1.16 M_{\odot}$, $T_{\rm eff}= 11\,650$ K and
$\log(M_{\rm H}/M_{\star})= -6$. In Table \ref{table:1} we show
the periods of the best-fit model along with the harmonic degree, the
radial order, and the period differences (theoretical minus
observed). Most of the periods of BPM~37093 are identified as $\ell= 2$
modes. This is not expected due to geometric cancellation effects
(that is, $\ell= 1$ modes should be more easily detectable than $\ell= 2$
modes). In Table \ref{table:2}, we list the main characteristics of the
best-fit model for BPM~37093. The parameters of the best fit model are in
agreement with the spectroscopically derived ones. On the other hand, 
the asteroseismological distance differs somewhat from the astrometric distance
obtained with {\it Gaia}.

\begin{acknowledgements}
  A.H.C. warmly thanks  the Local Organising Committee,
  in particular Prof. Werner W. Weiss,
  for support that allowed him to attend this conference.
\end{acknowledgements}

\bibliographystyle{aa}  
\bibliography{corsico_5p01} 

\end{document}